\documentclass[conference]{IEEEtran}
\usepackage[breaklinks]{hyperref}

\usepackage{cite}

\ifCLASSINFOpdf
   \usepackage[pdftex]{graphicx}
   \graphicspath{{../pdf/}{../jpeg/}}
   \DeclareGraphicsExtensions{.pdf,.jpeg,.png}
\else
   \usepackage[dvips]{graphicx}
   \graphicspath{{../eps/}}
   \DeclareGraphicsExtensions{.eps}
\fi
\usepackage{url}

\begin{document}
%
\title{Multi-core: Adding a New Dimension to Computing}

\author{\IEEEauthorblockN{Md. Tanvir Al Amin}
\IEEEauthorblockA{Department of Computer Science and Engineering\\
Bangladesh University of Engineering and Technology\\
Dhaka- 1000, Bangladesh\\
Email: tanviralamin@cse.buet.ac.bd}}

\maketitle

\begin{abstract}
Invention of Transistors in 1948 started a new era in technology, called Solid State Electronics.
Since then, sustaining development and advancement in electronics and fabrication techniques has caused the devices to shrink in size and become smaller, paving the quest for increasing density and clock speed. That quest has suddenly come to a halt due to fundamental bounds applied by physical laws. But, demand for more and more computational power is still prevalent in the computing world. As a result, the microprocessor industry has started exploring the technology along a different dimension. Speed of a single work unit (CPU) is no longer the concern, rather increasing the number of independent processor cores packed in a single package has become the new concern. Such processors are commonly known as multi-core processors.
Scaling the performance by using multiple cores has gained so much attention from the academia and the industry, that not only desktops, but also laptops, PDAs, cell phones and even embedded devices today contain these processors. In this paper, we explore state of the art technologies for multi-core processors and existing software tools to support parallelism. We also discuss present and future trend of research in this field. From our survey, we conclude that next few decades are going to be marked by the success of this ``Ubiquitous parallel processing''.
\end{abstract}


%
\IEEEpeerreviewmaketitle

\section{Introduction}
In 1965, Intel co-founder Gordon Moore predicted ~\cite{moore} that the transistor density of semiconductor chips would double roughly every 18 months, which is commonly known as Moore's Law. Shrinking the feature size by a factor of $x$ means clock rate can be increased nearly by a factor of $x$, and number of transistors per unit area can go up by a factor of $x^2$. Due to sustaining advancement in VLSI fabrication technology, the computing industry had been experiencing this evolution of computational power through clock speed, until a wall was hit due to fundamental physics. As fabrication density increases, manufacturing costs go up, and more importantly, power density becomes higher. Moreover, though computing power increases linearly with clock speed, heat dissipation rises with the square or cube, depending on the electrical model, and clock frequencies beyond 5 GHz easily melt chips ~\cite{survive}. Thus the ``free lunch'' of performance -- in the form of ever faster clock speeds -- is over.

This physical boundary has forced microprocessor vendors to take a new route -- increase performance through parallelism. It is now commonly accepted that as serial processing has reached a technology edge, it is parallel processing which can save the day, if applied correctly ~\cite{berkeleyview}. Parallel processing was once the topic of mainframe or grid processing powered data centers only, but now it is the story of mass consumer market, as we have so called ``Multi-core Processors'' -- two or more independent processor cores in a single package ~\cite{wiki}.

Multi-core processors, though leverage the power of parallel processing, are different from Multiprocessor systems of supercomputers or mainframes, both in technical and application framework. Multiprocessor systems have either uniform or non uniform memory access, and processor to processor communication implies more loss in signal, as off-chip delay is far greater than on-chip delay ~\cite{wiki}. On the other hand, multi-core systems work in shared memory, and in a lot of cases shared cache systems. Processor to processor communication is not expensive as both are in a single package.

Amdahl's law limits the speedup from using parallelism ~\cite{survive}. If a fraction $p$ of a computation can be run in parallel, and the rest must run serially, maximum speedup is $\frac{1}{1-p}$. This law applies for data centers or batch processing jobs, but the scenario with multi-core processors is a bit more optimistic, as consumer level computers, due to multitasking, has inherent parallelism of threads. This environment ensure the performance boast through multi-core, because even if softwares are not written specifically for multi-core, in a multi-threaded way; there are already lots of threads running in a personal computer. Use of multi-core processors easily improves the response time and perceived performance here.

In this paper, we discuss existing technologies for multi-core processors. Almost all microprocessor manufacturing companies have produced multi-core versions of their product. Dual-core or Quad-core processors from Intel or AMD are available in market for desktop, laptop or small servers \cite{wiki}. Latest graphics processors (GPU) from NVidia or ATI (AMD) are also multi-core.
Broadcom, Cavium or TI are already producing multi-core embedded processors for various DSP or communication necessities ~\cite{wiki}.
Even the gaming consoles are not apart from this multi-core wave. We also discuss ``RAMP'', an approach from UC-Berkeley ~\cite{berkeleyview}, low clock-speed manycore processors on FPGA boards, designed as a testbed for scalability of multi-core applications.

As multi-core operation is another form of parallel computing, scalability depends on how much parallel the tasks are, especially when we have lots of cores. Thus software redesign exploiting parallelism and multithreading with load balancing is necessary for ultimate performance gain. Already several tools have been developed for this ~\cite{survive}. In this paper, we discuss CILK++, NI LabView, Intel TBB and several other frameworks.

We observe that the industry is approaching a transition from multi-core to manycore. Intel has already started its Tera-scale computing research. Tilera released TILE64, a 64 core processor. A new form of Moore's law is now in effect, \emph{``The number of cores in a physical processor will double in every technology generation''.}

\section{Parallel Computing Revolution}

\subsection{ILP and TLP}
Chip designers were always in the quest of increasing throughput of a processor. First and obvious approach is to divide the instruction in multiple steps and use pipelining ~\cite{patthenn}. As multiple instructions are executed concurrently, this is termed as ILP (Instruction Level Parallelism). ILP can be increased more by making the pipeline stages multiple issue, i.e. replicating some of the internal components and launching multiple instructions in every pipeline stage. Techniques like loop unrolling, superscalar, dynamic prediction, register renaming were \cite{patthenn} devised to find independent instructions. But these techniques proved inefficient for hard to predict codes and it seemed that parallelism could be exercised better, if independent execution paths were already specified in the program code ~\cite{wiki}. This resulted in the idea of TLP (Thread Level Parallelism) i.e. multiple thread-multiple core.

\subsection{Multi-core Processors}
\begin{figure}
\begin{center}
\begin{tabular}{c}
\resizebox{50mm}{!}{\includegraphics{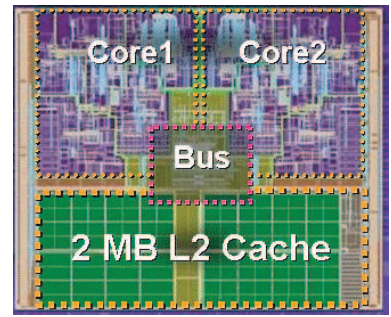}}\\
\end{tabular}
\caption{An example configuration of a dual-core processor
        }
\label{topology}
\end{center}
\end{figure}

A multi-core processor contains two or more independent processing cores in a single package. A dual-core processor contains two cores in a single die (Fig. ~\ref{topology}), and a quad-core processor contains four cores in one or two dies.

Each ``core'' independently implements optimizations such as superscalar execution, pipelining, and multithreading. The processors share the same interconnection to the rest of the system. Hence the bus and memory is shared in general. This is a difference from typical multiprocessing, where each processor in general have separate memory. But, as each core are independent, they can effectively run separate threads in parallel. Thus a multi-core microprocessor is an implementation of multiprocessing in a single physical package, and it is most efficient when each are presented with independent parallel tasks. These CPUs either use homogeneous (each processor similar in power and responsibility) or heterogeneous architecture (hierarchical
processor arrangement) ~\cite{wiki}.

Whether the cores share a single coherent cache at the highest on-device cache level or may have separate caches is largely implementation and application dependent. Number of cores fabricated on a die is also design dependent. A processor with all cores on a single die is called a monolithic processor.

\section{Present multi-core Processors}
\subsection {IBM}
IBM POWER4 chip released in 2001 implemented 64-bit PowerPC ~\cite{power4} architecture.
It is the first commercial multi-core microprocessor, with two cores on a single die. 174 million transistors were fabricated using a 180 $nm$ CMOS 8S3 SOI process.  It consumed 115W power with a Vdd of 1.6V.
\subsection{Intel}
\begin{figure}
\begin{center}
\begin{tabular}{c}
\resizebox{70mm}{!}{\includegraphics{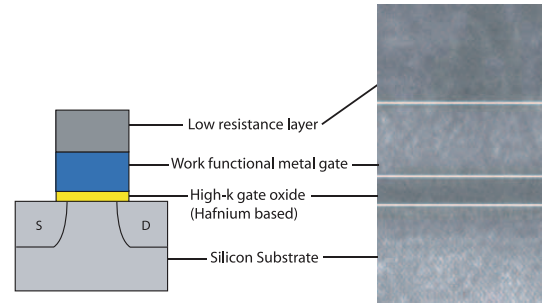}}\\
\end{tabular}
\caption{The high-k Transistors
        }
\label{highkt}
\end{center}
\end{figure}

\begin{figure}
\begin{center}
\begin{tabular}{c}
\resizebox{60mm}{!}{\includegraphics{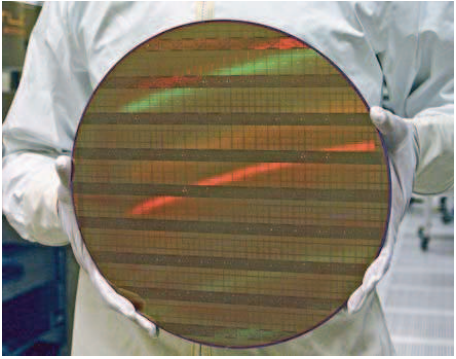}}\\
\end{tabular}
\caption{45 nm process wafers using high-k transistors
        }
\label{highkw}
\end{center}
\end{figure}

Dual core Intel processors based on NetBurst microarchitecture (Pentium D) were not successful due to high power consumption and cooling problems. First successful Intel multi-core processors were based on Intel Core microarchitecture ~\cite{coreduo} , which was actually targeted for mobile platform. Core 2 microarchitecture was a revision of Core, and several processors like Core 2 Duo, Core 2 Quad or Core 2 Extreme for desktop pc's were released. Core 2 Duo contained 2 processors on a single die, and Core 2 Quad contained 2 such dies in a single package. Before 2007, these processors were fabricated using a 65 $nm$ process. However, Intel developed a new technology MOS transistor replacing ordinary $SiO_2$ insulator by Hafnium high-k gate oxide (Fig ~\ref{highkt}), which scaled the process down to 45 $nm$ ~\cite{highk}. Newer multi-core processors based on this process, codenamed ``Penryn'' (Fig ~\ref{highkw}) define the state of the art technology for Intel.

\begin{figure}
\begin{center}
\begin{tabular}{c}
\resizebox{60mm}{!}{\includegraphics{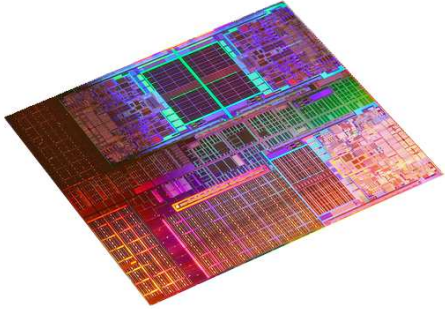}}\\
\end{tabular}
\caption{Die micrograph of Intel's Dunnington hexa-core processor
        }
\label{dunning}
\end{center}
\end{figure}
Multi-core Xeon CPU's for server line computers include several processors including Clovertown, Harpertown and latest Dunnington. Dunnington is a single die hexa-core processor (Fig ~\ref{dunning}), with 3 unified 3MB L2 caches, 96KB L1 cache and 16 MB L3 cache. TDP is less than 130W.

\subsection{AMD}

\begin{figure}
\begin{center}
\begin{tabular}{c}
\resizebox{50mm}{!}{\includegraphics{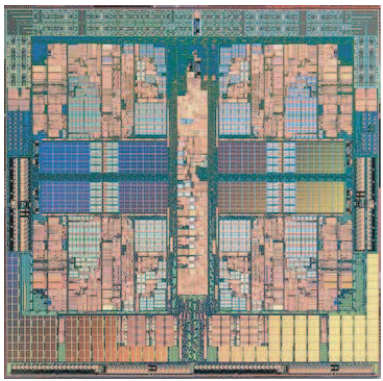}}\\
\end{tabular}
\caption{AMD quad-core Barcelona die shot
        }
\label{amdbarcelona}
\end{center}
\end{figure}
Like Intel, AMD also produces similar chips named Athlon 64 X2, Opteron, Phenom and so on. Their quad-core processor is
codenamed barcelona (Fig. ~\ref{amdbarcelona}). AMD, after acquisition of ATI, also produces multi-core stream processors. FireStream is such a processor with 10 cores, having 16 5-issue wide superscalar stream processor per core.

\subsection{Gaming Consoles}
\begin{figure}
\begin{center}
\begin{tabular}{c}
\resizebox{60mm}{!}{\includegraphics{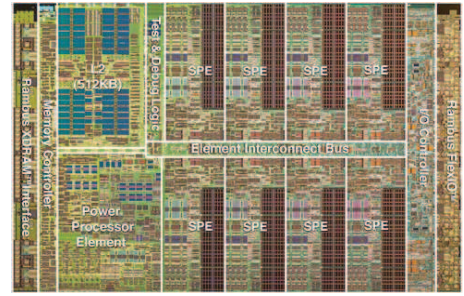}}\\
\end{tabular}
\caption{PlayStation3 Cell octa-core Processor
        }
\label{cell}
\end{center}
\end{figure}

PlayStation3 features Cell processor. ``Cell Broadband Engine'' Architecture is jointly developed by Sony, Toshiba and IBM. Cell is a octa-core processor having novel memory coherence architecture. (Fig ~\ref{cell})
The architecture emphasizes efficiency/watt, prioritizes bandwidth over latency, and favors peak computational throughput over simplicity of program code.

Xbox 360 features another multi-core processor named Xenon. This processor is based on IBM's PowerPC instruction set architecture, consisting of three independent cores on a single die. Each of the cores has two symmetric hardware threads (SMT), for a total of six hardware threads available to games. Each individual core also includes 32 KB of L1 instruction cache and 32 KB of L1 data cache.

\subsection{NVidia}
NVidia produces GPU of GeForce series and GPGPU (General purpose GPU) of Tesla series (CUDA framework)
These processors and latest generation 9 or generation 10 GPU's are multi-core as well.

Several other manufacturers like Sun, ARM, Cavium, Broadcom, Infineon also has their own multi-core processors.
Some of these are solely for embedded applications and some are for specialized communication or DSP.

\subsection{RAMP}
\begin{figure}
\begin{center}
\begin{tabular}{c}
\resizebox{55mm}{!}{\includegraphics{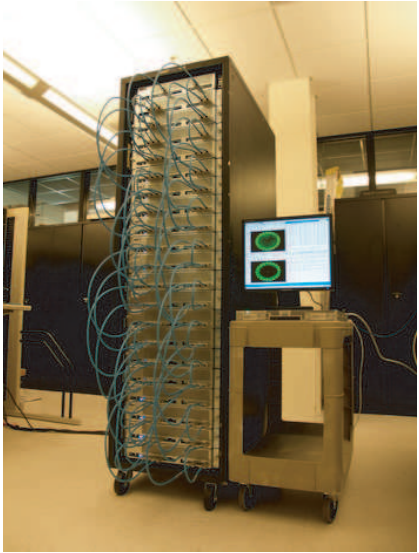}}\\
\end{tabular}
\caption{RAMP Blue V3.0 1008 Core, Rack and Server
        }
\label{ramp}
\end{center}
\end{figure}

RAMP (Research Accelerator for Multiple Processors) is an approach for evaluating newer hardware and software architectures
on multi-core processors. It can be termed as ``Academic Manycore'' built on FPGA boards ~\cite{berkeleyview}. RAMP is a real
world simulator for large manycore systems, i.e it uses soft-core processors with entire logic built on FPGA boards.
Presently they run at 90-150 MHz speed and are capable of executing real applications in unmodified binary. RAMP processors are created with 256, 768 or 1008 soft core processors and are ideal for heterogeneous chip architectures. Fig ~\ref{ramp} shows one such processor RAMP Blue V3.0. The research group
at Berkeley describe RAMP as \emph{a vehicle to lure more researchers to parallel challenge and decrease time to parallel salvation.}

\section{Software Architecture}
General vision for multi-core based software is Scalability, Code simplicity and Modularity ~\cite{survive}.
To exploit the processing power offered by multi-core, potential opportunity for multithreading and load balancing must be recognized by the
programmer. Not only that, writing efficient parallel and multithreaded programs also need careful reasoning, because there are
now provisions for hard to track bugs created by wrong synchronization, race condition or deadlock. Several concurrency platforms are available today.

LabVIEW (Laboratory Virtual Instrumentation Engineering Workbench) from National Instruments features a graphical language, named ``G'' which is a dataflow programming language and is inherently capable of parallel execution.

MPI Message passing was the classical method of choice for the HPC (High Performance Computing) committee.
Scientific codes written with MPI have been ported to multi-core systems and they run quite well.

Intel Threading Building Blocks (TBB) is an open-source C++ template library containing data structures and algorithms  for writing task-based multithreaded applications.

OpenMP is an open-source concurrency platform supporting multithreaded programming through Fortran and C/C++ language \emph{\#pragma}.

CILK++ from Cilk Arts, extends C++ to multi-core via three keywords cilk\_spawn, cilk\_sync and cilk\_for (parallel loops). Cilkscreen race detector can be used to test a Cilk++
binary program for its parallel correctness.

\section{Future of multi-core Computing}
It is now widely agreed that multi-core is going to be the wave of the future.
But there are differing views ~\cite{eetimes} about processor organization, hierarchy and resulting software model of the future.
One group believe that the cores should be heterogeneous in nature, using specialty cores for different jobs, scheduled by high level API.
On the other hand, the other group opposes this idea and expresses that, multi-core chips need to be homogenous
collections of general-purpose cores to keep the software model simple.

\begin{figure}
\begin{center}
\begin{tabular}{c}
\resizebox{70mm}{!}{\includegraphics{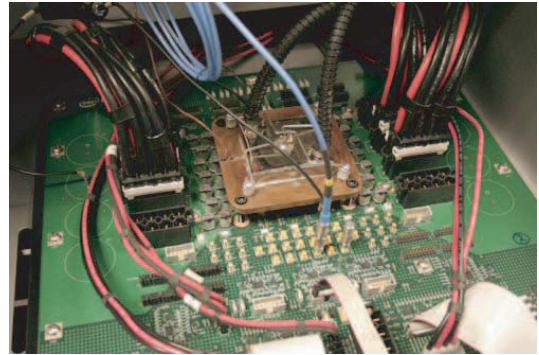}}\\
\end{tabular}
\caption{Teraflops Research Chip as demonstrated in 2007
        }
\label{80core}
\end{center}
\end{figure}

Intel ~\cite{fewcore2many} Tera scale research program aims to scale today's multi-core architectures
to 100s of cores. Their vision is to create platform capable of performing Tera flops of operation on Tera bytes of data. The Teraflops Research Chip (also called Polaris) is the first processor prototype developed by Intel's Tera-scale Computing Research Program. A brief demo was presented the IDF on September 2006 and a working model was shown at the 2007 ISSCC. (Fig ~\ref{80core}).

The chip contains 80 cores, constructed using a 65 nm CMOS process \cite{teraflop}. Each core contains two programmable floating point engines and one 5-port messaging passing router. Running at 3.16 GHz it achieved 1.01 TFLOPS \cite{inquirer} with a total power consumption of 62 W and a on-chip temperature of 383 K. Increasing frequency to 5.7 GHz and power to 265 W increased performance to 1.81 TFLOPS and later 2 TFLOPS.

Multi-core processors will continue to dominate at least until some new device theory offering new limits for density, power efficiency and limit of computational speed is available.
Present barriers imposed by complex software requirement will be resolved as more and more research initiatives will be taken on parallel and multi-core programming.
Hence we can conclude that, by doubling the number of cores every technology generation, soon we are going to enter the realm of Tera Scale Computing.

\bibliographystyle{IEEEtran}
\bibliography{mybib}

\end{document}